\documentclass[12pt]{article}
\input{epsf}
\usepackage{psfig,epsf}
\usepackage{amsmath}

\newcommand{\be}{\begin{equation}}
\newcommand{\ee}{\end{equation}}
\newcommand{\ba}{\begin{eqnarray}}
\newcommand{\ea}{\end{eqnarray}}

\begin{document}

\title{Production of $\Theta^+$ Hypernuclei with the $(K^+,\pi^+)$
reaction}

\author{H. Nagahiro$^a$, S. Hirenzaki$^a$, E. Oset$^b$ and M.J. Vicente
Vacas$^b$\\
{\small $^a$Department of Physics, Nara Women's University, Nara
630-8506, Japan} \\
{\small $^b$Departamento de F\'{\i}sica Te\'orica and IFIC,
Centro Mixto Universidad de Valencia-CSIC,} \\
{\small Institutos de
Investigaci\'on de Paterna, Aptd. 22085, 46071 Valencia, Spain}\\
}

\date{\today}

\maketitle
\begin{abstract}

We present results on the production of bound states of $\Theta^+$ in
nuclei
using the $(K^+,\pi^+)$ reaction.  By taking into account the  states
obtained
within a wide range of strength of the $\Theta^+$ nucleus optical
potential,
plus the possibility to replace different nucleons of the nucleus, we
obtain an
excitation spectra with clearly differentiated  peaks. The magnitude of
the
calculated cross sections is well within  reachable range.
\end{abstract}

The discovery of the $\Theta^+$ at SPring-8/Osaka \cite{nakano},
followed by its
confirmation in different other experiments, has made a substantial
impact in
hadronic physics (see \cite{hyodo} for a compilation of experimental 
and
theoretical works on the issue).
The possibility that there would be $\Theta^+$  bound states in nuclei
has not
passed unnoticed and in \cite{kim} the $\Theta^+$ selfenergy in the
nucleus was
evaluated, however with only the part tied to the $K N$ decay, which is
known
experimentally to be very small. As a consequence, the $\Theta^+$
potential obtained was too weak to bind $\Theta^+$ in nuclei.
Suggestions of
possible bound states within a schematic model for quark pair 
interaction with
nucleons were made in Ref. \cite{miller}. In that work a $\Theta^+$ 
selfenergy
in the nucleus was obtained of the same order as  the contribution of 
the same
mechanisms to the $\Theta^+$ mass. The
$(\gamma,\Sigma)$ and $(K,\pi)$ reactions were also suggested in that 
work
as a means to produce $\Theta^+$ bound states in nuclei.

    In a recent paper \cite{hyper} the possibility of having 
$\Theta^+$  bound
states in nuclei,  tied to the $K\pi N$ content of the $\Theta^+$, was
investigated and it was concluded that there  is an attractive 
$\Theta^+$
potential, which, within uncertainties,  is strong  enough to bind the
$\Theta^+$ in nuclei. Restrictions from Pauli blocking and binding 
reduce the
$\Theta^+$ width  in nuclei to  about one third or less of the free 
width, and
with attractive $\Theta^+$  nucleus  potentials ranging from 60 to 120 
MeV at
normal nuclear matter density,  the separation between the deeper 
$\Theta^+$
levels in light and medium nuclei is larger than the  width, even in 
the case
that the free $\Theta^+$ width were as big as 15 MeV.  This is a 
desirable
experimental situation in which clear peaks could be observed provided 
an
appropriate  reaction is used.

    In \cite{hyper} the  $\Theta^+$ selfenergy tied to the $K N$ decay
was also studied and found to be very small like in \cite{kim}. The 
large
attraction found in \cite{hyper} is tied to
the coupling of the $\Theta^+$ to two mesons and a baryon
which was related to
the strong
decay of the $N^*(1710)$ resonance to a nucleon and two pions.

In the present paper we investigate the  reaction $(K^+,\pi^+)$ in  
nuclei,
which leads to clear peaks and a fair strength in the spectra of  
$\Theta^+$
nuclear states. The reaction is of the substitutional type, in which 
one of
the  nucleons in the nucleus will be substituted by the $\Theta^+$.  
Hence, on
top of the  different $\Theta^+$ bound states, we shall also have to 
take into
account the  binding energies of the nucleon levels in the nucleus 
when we look
for the  spectra of this reaction.

The analogous $(K^-,\pi^-)$ reaction has been a standard tool to  
produce
$\Lambda$ hypernuclei \cite{hyperlam} and it was also used to see the 
only
$\Sigma$ hypernucleus known so far \cite{hypersig,hypersig2}.   There 
is an
added difficulty here since one would wish to have a kinematics as 
close as
possible to a recoilless condition, which makes the production cross 
sections
larger. This is the case of the $(d,^3 He)$ reaction, which was used 
with
success to  find the deeply bound pionic atoms 
\cite{yamazaki,hirenzaki}. In
the present case,  given  the fact that one has to create the excess 
mass of
the $\Theta^+$ over the  nucleon from the kinetic energy of the kaon, 
this
induces inevitably a  momentum transfer, which reach a minimal value  
of the
order of 450 MeV/c  for an initial $K^+$ momentum of around 620 MeV/c. 
Although
this quantity is larger than the Fermi momentum, it can still  be 
accommodated
in the  $\Theta^+$ wave function without much penalty,  such that the 
cross
sections obtained are still sizable.

\begin{figure}[hbt]
%Fig.(K^+,\pi^+) momentum transfer
\epsfxsize=12cm
\centerline{
\epsfbox{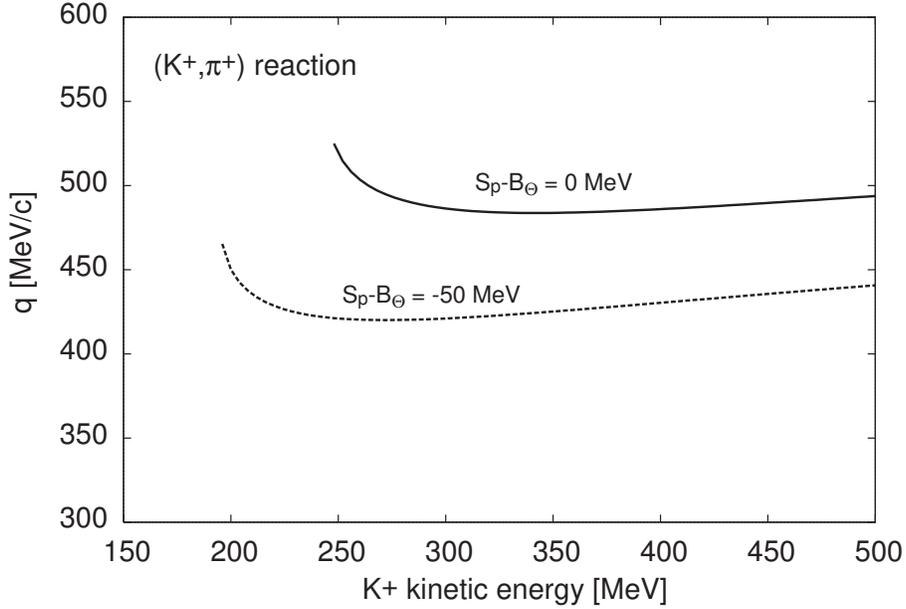}
%\fbox{\rule{11cm}{0cm}\rule{0cm}{5cm}}
}
\caption{
Momentum transfer of the ($K^+$,$\pi^+$) reaction for the formation of
 the $\Theta^+$ nuclear states plotted as a function of the incident
 kaon kinetic energy. The solid line shows the result with 
$S_p-B_\Theta=0$
 and the dashed line with $S_p-B_\Theta=-50$ MeV, where $S_p$ and
 $B_\Theta$ are the proton separation energy and $\Theta$ binding
 energy, respectively.
\label{fig:mom_trans}
}
\end{figure}

 We show the momentum transfer of the
forward ($K^+$,$\pi^+$) reactions for the $\Theta^+$ nuclei formation 
as a
function of the incident $K$  energy in Fig.\ref{fig:mom_trans}. We 
can also
expect to produce $\Theta^+$ nuclear states by the ($\gamma$,$K^-$) and
($\pi^-$, $K^-$) reactions. However, the typical momentum transfer of 
these
reactions are much larger than that of the ($K^+$,$\pi^+$) reaction, 
and the
formation rate of these reactions will be suppressed significantly.  
Also the
distortion of the $K^+$ is substantially weaker than that of the  
$K^-$.

The cross section for the ($K^+$,$\pi^+$) reaction
for the formation of the $\Theta^+$ bound states
is given by,
\begin{equation}
\frac{d\sigma}{d\Omega_\pi d\omega_\pi} =
\frac{1}{16\pi^2}\frac{p}{k} |t|^2
\sum_{
\begin{subarray}{c}
{\rm p-state},\\
{\rm \Theta-state}
\end{subarray}
}
\frac{\Gamma}{2\pi}\frac{1}{(\Delta E)^2+\Gamma^2/4}
\left| \int d^3\vec{r}
\psi_\Theta^\dagger(\vec{r}) e^{i(\vec{k}-\vec{p})\cdot\vec{r}}
D(\vec{b},z)\psi_p(\vec{r})
\right|^2,
\label{eq:d_sigma}
\end{equation}
where $\psi_\Theta(\vec{r})$ and $\psi_p(\vec{r})$ are the $\Theta$ and
proton wavefunctions in bound states, $\vec{k}$ and $\vec{p}$ are the
momenta of the incident kaon and the emitted pion, respectively, and
$\Gamma$ is the decay width of the final $\Theta^+$ nuclear states.
In Eq.~(\ref{eq:d_sigma}) $\Delta E$ is defined as,
\begin{equation}
\Delta E = \omega(K^+) + (M_p-S_p)-(M_\Theta-B_\Theta)-\omega(\pi^+),
\end{equation}
where $\omega(K^+)$ and $\omega(\pi^+)$ are the relativistic energies 
of
$K^+$ and $\pi^+$, $S_p$ is the proton separation energy from the 
target
nucleus, $B_\Theta$ is the $\Theta$ binding energy in the final
states and
$M_p$, $M_\Theta$ are the proton and $\Theta$ masses,
respectively.
We sum up all contributions of proton single particle states in the 
target
nucleus and $\Theta$ bound states in the final states in
Eq.~(\ref{eq:d_sigma}).

In order to take into account the distortion affecting the pion and
kaon, we use the eikonal approximation, in which the distorted waves 
are
approximated by plane waves with a distortion factor. The distortion
factor $D(\vec{b},z)$ appearing in Eq.~(\ref{eq:d_sigma}) is defined 
as,
\begin{equation}
D(\vec{b},z)=\exp\left[
-\frac{1}{2}i\int_{-\infty}^z
\frac{\Pi_K}{k} dz'
-
\frac{1}{2}i\int_z^\infty \frac{\Pi_\pi}{p} dz'
\right],
\label{eq:distortion}
\end{equation}
where $\Pi_K$ and $\Pi_\pi$ are the  kaon and
pion selfenergies in the nuclear medium. The real part of $\Pi_K$ is 
taken
from the $t\rho$ approximation  \cite{Kaiser:1996js,Oset:2000eg}
\begin{equation}
Re \,\Pi_K=0.13 \, m_K^2 \rho/\rho_0,
\end{equation}
which accounts for the largely dominant $s-$wave interaction and the 
imaginary
part, also from the $t\rho$ approximation, is obtained using the 
optical theorem
\begin{equation}
\frac{Im\, \Pi_K}{k}=-(\sigma_{Kp} \rho_p + \sigma_{Kn} \rho_n)\, .
\end{equation}
The pion selfenergy $\Pi_\pi$ is taken from Refs.~\cite{NPA454,NPA554,
NPA705}.

The matrix element $t$ in Eq.~(\ref{eq:d_sigma}) is the
$K^+ p \to \pi^+ \Theta^+$ transition  $t-matrix$.  We get this
magnitude from the same Lagrangian used in \cite{hyper} to
obtain the
$\Theta^+$ nuclear potential.
The $\Theta^+$ selfenergy in \cite{hyper} was obtained studying the
excitation of intermediate states $KN$ and $K \pi N$, and incorporating
the medium effects in the mesons and the nucleon.  The part of the
selfenergy tied to the $KN$ channel was found very small, but the one
tied to the $K \pi N$ channel led to a sizable attraction.

In Ref. \cite{hyper} the following two Lagrangians were used coupling
the $\Theta^+$ to $K \pi N$
\begin{equation}
 {\cal L} = i g_{\bar{10}} \epsilon ^{ilm} \bar{T} _{ijk} \gamma^{\mu}
 B^j_l (V_{\mu})^k_m ,
\end{equation}
\begin{equation}
 {\cal L} = \frac{1}{2f} \tilde{g}_{\bar{10}} \epsilon ^{ilm} \bar{T}
_{ijk}
 (\phi \cdot \phi )^j_l B^k_m ,
\end{equation}
with $V_{\mu}$ the two mesons vector current,
$V_{\mu}=\frac{1}{4f^2} (\phi\partial_{\mu}\phi- 
\partial_{\mu}\phi\phi)$,
 $f$ the pion decay constant
and $T _{ijk}$, $B^j_l$, $\phi^j_l$ SU(3) tensors which account for the
antidecuplet states, the
octet of $\frac{1}{2} ^+$ baryons and the octet of $0^-$ mesons,
respectively.

In Ref.\cite{Hosaka:2004mv} a study was done of the possible  
Lagrangians
coupling the antidecuplet states to two meson and one baryon under the
assumptions of $SU(3)$ symmetry and minimal number of derivatives for 
each
possible $SU(3)$ structure. In addition, a possible chiral symmetric 
term was
also studied as well as a mass term breaking explicitly chiral 
symmetry.
It was found there that the two relevant structures were those in Eqs. 
(6) and
(7). The chiral Lagrangian gave results remarkably similar to those of 
Eq. (6)
and other terms as the chiral symmetry breaking term and one coming 
from the
27 $SU(3)$  representation had to have small strength from physical 
grounds.
 The amplitude provided by the first Lagrangian,
in the
nonrelativistic
limit, is proportional to the difference of the meson energies in the
$\Theta^+ \to N
\pi K$ process.  Here, since we have an incoming and an outgoing meson,
the
difference of energies is substituted by their sum.
Hence, the $ K^+ p \rightarrow \Theta^+ \pi^+$ amplitude is written as,
\begin{equation}
t=-\frac{1}{4f^2}(-\sqrt{6})g_{\bar{10}}(\omega(K^+)+\omega(\pi^+)),
\end{equation}
where $f=93$ MeV and $g_{\bar{10}}=\alpha \left[
\frac{m_{K^*}^2}{m_{K^*}^2-(k-p)^2}\right]$ with $\alpha=0.315$
\cite{Hosaka:2004mv}, which incorporates a form factor to explicitly 
account for
the exchange of a $ K^*(892)$ in the $t-$channel.

The amplitude provided by the second Lagrangian, Eq. (7), is
\begin{equation}
t=\frac{1}{2f}\sqrt{6}\tilde{g}_{\bar{10}},
\end{equation}
with $\tilde{g}_{\bar{10}}=1.88$.

The strength of these two terms is similar and they interfere, but 
their
relative phase is unknown. That phase is not relevant for the 
evaluation of the
selfenergies in Ref. \cite{Hosaka:2004mv} since there one studies the 
$\pi K$
production where the meson pair comes in $p-$wave and $s-$wave from the
Lagrangians of Eqs. (6) and (7) respectively.

Hence, in the present reaction we must admit a quite large 
uncertainty. There is
a hint on which relative sign  to take, based on the small upper bound 
for
$\Theta^+$ production in the preliminary results of K. Imai et 
al.\cite{imai}
for the reaction $(\pi^-,K^-)$ on the proton. This small cross section 
could be
qualitatively understood in base to a negative interference of the two 
terms
that we have. For the case of $(K^+,\pi^+)$ the sign of the amplitude 
for the
vector Lagrangian of Eq. (6) is opposite to that of the $(\pi^-,K^-)$ 
reaction,
while the amplitudes from Eq. (7) do not change. Based upon this, we 
take
the positive relative sign between the amplitudes.

   The wave functions of the $\Theta^+$ are obtained by solving the
Schr\"{o}dinger
equation with two potentials, one with a strength
\begin{equation}
V(r)=-60\frac{\rho(r)}{\rho_0} {\rm [MeV]},
\end{equation}
and the other one
\begin{equation}
V(r)=-120\frac{\rho(r)}{\rho_0} {\rm [MeV]}.
\end{equation}
The calculated binding energies with these potentials are reported in
Ref.~\cite{hyper}. In this exploratory level, we only take into account
a volume type potential and we have no $LS$ splitting in the $\Theta$
energy
spectrum.
As to the width, it was found in \cite{hyper} that assuming the free
width to be
15 MeV the width in the medium was smaller than 6-7
MeV, due
to Pauli blocking effects mostly.
We evaluate the level widths using the imaginary part of the
$\Theta$ selfenergies calculated in Ref.~\cite{hyper} at the 
appropriate
$\Theta$
energies.

With this range of values we intend to account for different 
uncertainties
discussed in Ref.\cite{hyper}, like the experimental uncertainties in 
the input
used to fix the  $\tilde{g}_{\bar{10}}$ and ${g}_{\bar{10}}$ couplings,
uncertainties in the nucleon selfenergies, etc.

\begin{figure}[hbt]
%Fig.12C(K,pi) V=60 ver. (graph)
\epsfxsize=12cm
\centerline{
%\epsfbox{12CV60graph.eps}
\epsfbox{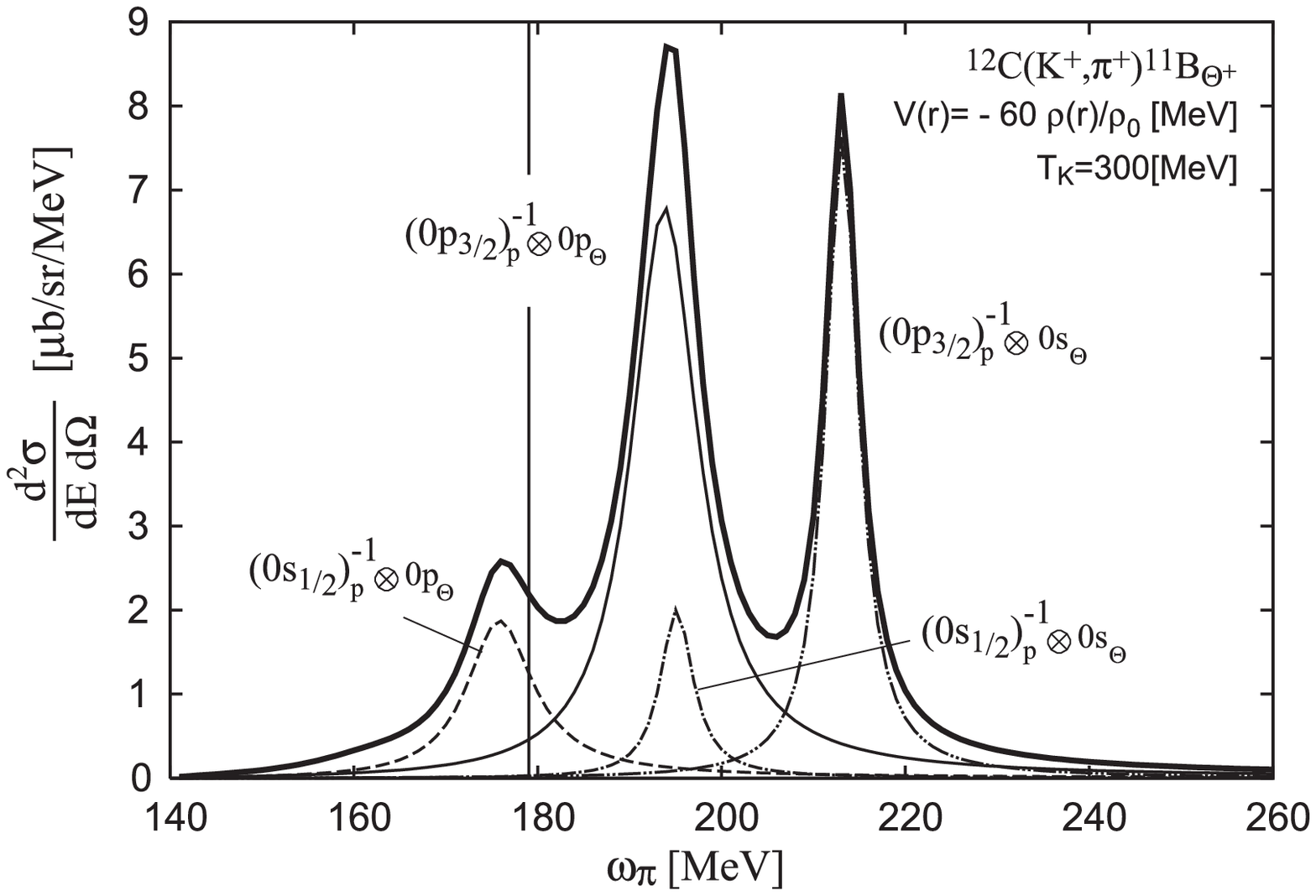}
%\fbox{\rule{11cm}{0cm}\rule{0cm}{5cm}}
}
\caption{
Calculated $\Theta$ bound states formation cross section
shown as a function of the emitted pion
energy $\omega_\pi$
at forward angles
for a $^{12}$C target.
The incident kaon kinetic energy, $T_K$, is 300 MeV, and the shallow
 $\Theta$ nuclear potential $V(r)=-60\rho(r)/\rho_0$ MeV is used.
The total spectrum is shown by the thick-solid line and the dominant
 subcomponents are also shown by the thin lines as indicated in the 
figures.
The $\Theta$ production threshold leaving the residual nucleus in its
ground state is shown by the vertical line.
\label{fig:12CV60}
}
\end{figure}

\begin{figure}[hbt]
%Fig.12C(K,pi) V=120 ver. (graph)
\epsfxsize=12cm
\centerline{
%\epsfbox{12CV120graph.eps}
\epsfbox{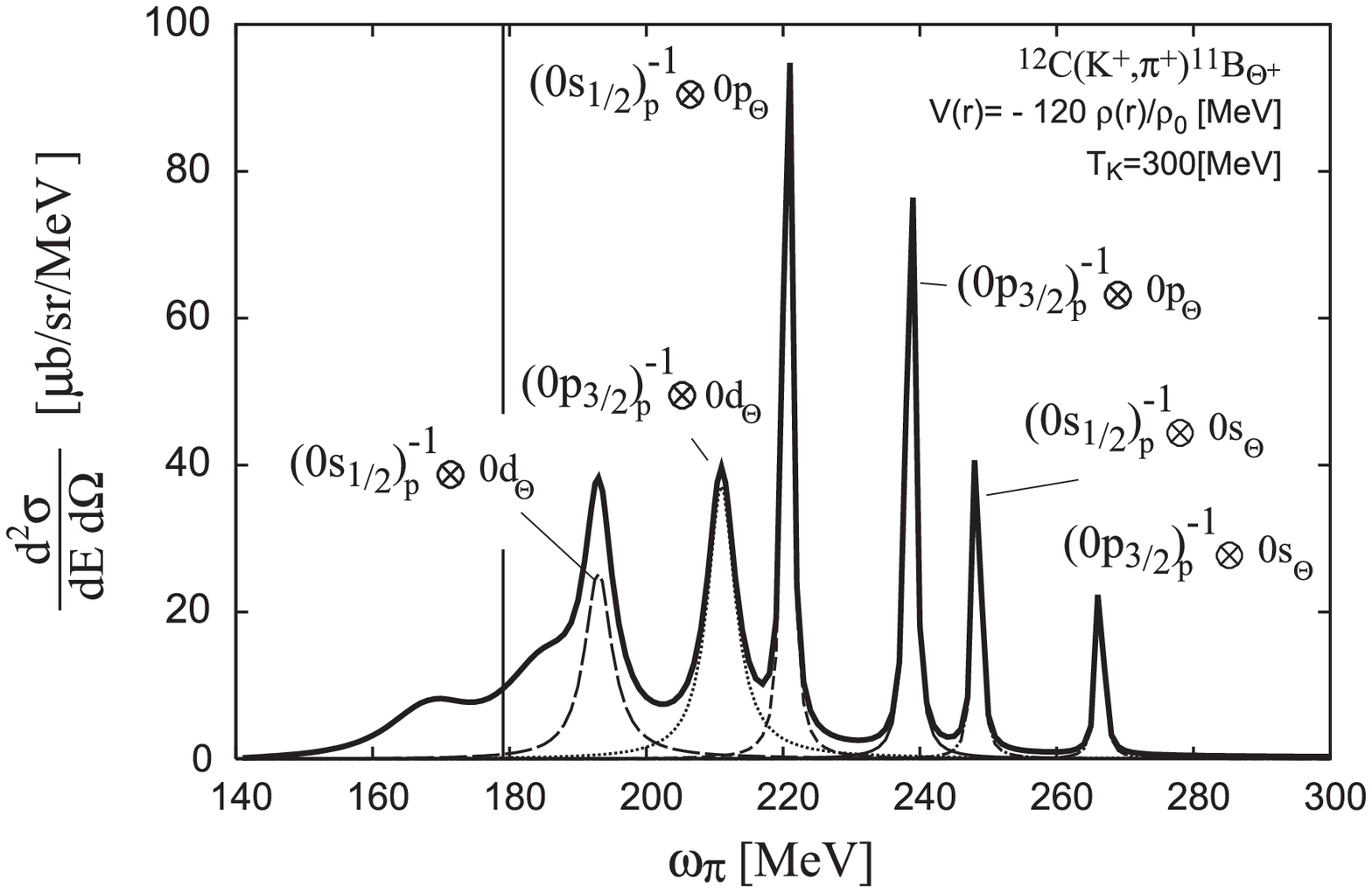}
%\fbox{\rule{11cm}{0cm}\rule{0cm}{5cm}}
}
\caption{
Calculated $\Theta$ bound states formation cross section
shown as a function of the emitted pion
energy $\omega_\pi$
at forward angles
for a $^{12}$C target.
The incident kaon kinetic energy, $T_K$, is 300 MeV, and the deep
 $\Theta$ nuclear potential $V(r)=-120\rho(r)/\rho_0$ MeV is used.
The total spectrum is shown by the thick-solid line and the dominant
 subcomponents are also shown by the thin lines as indicated in the 
figure.
The $\Theta$ production threshold leaving the residual nucleus in its
ground state is shown by the vertical line.
\label{fig:12CV120}
}
\end{figure}

%(b)
We show the calculated spectra for the formation of the $\Theta^+$ 
bound
states in Fig.~\ref{fig:12CV60} and Fig.~\ref{fig:12CV120}. We selected
carbon as  target since the level spacing of  each
subcomponent is expected to be comfortably large to observe the 
isolated
peaks structure. The incident kinetic energy of the kaon beam is fixed 
at 300 MeV to minimize the momentum transfer of the reaction.

The results with the shallow $\Theta$ potential $V(r)=-60\rho/\rho_0$
(MeV) are shown in Fig.~\ref{fig:12CV60}. We find three isolated
peaks in the spectrum. We also find that the
magnitude of the formation cross section is around a few [$\mu$b/sr 
MeV]
which is expected to be reachable in experiments.
The results with the deep $\Theta$ potential $V(r)=-120\rho(r)/\rho_0$
(MeV) are shown in Fig.~\ref{fig:12CV120}, where we find the separated
peaks in the cross section again.
In this case, we have six clear peaks in the spectrum according to the
existence of more $\Theta$ bound states due to the deeper potential.
The magnitude of the cross section is around 10 times larger than
with the shallow potential case because the width of the $\Theta$ 
state is
smaller for the deeper bound states which makes the peaks higher. The
number of subcomponents of the spectrum is increased for the case of 
the
deeper potential
and we find again reasonably large cross section
for the reaction.
We should mention here that the real parts of the distortion potentials
in Eq.~(\ref{eq:distortion}) are relevant and make the cross sections
larger by about a factor two for both $\Theta$ potential cases.

In these calculated results, we have not included the
quasi-free $\Theta$ production, which will have certain contribution to
the spectrum above the $\Theta$ production threshold. The threshold is
shown by the vertical lines in Fig.~\ref{fig:12CV60} and
\ref{fig:12CV120}. The spectrum for $\omega_\pi$ below the
threshold would be modified by the inclusion of the quasi-free $\Theta$
processes.
However, the spectrum in the bound $\Theta$ region will not be affected
by them.

In summary, the  results of \cite{hyper}  indicate that there should be
bound
states of $\Theta^+$  in nuclei, with separation energies reasonably
larger
than the width of the states. In view of that, we investigated the
$(K^+,\pi^+)$ reaction to produce these states and obtained excitation
spectra
of  $\Theta^+$ states for a $^{12}$C target
with two different
potentials
which cover the likely range of the $\Theta^+$ nucleus optical
potential according
to the calculations of \cite{hyper}.
With the caveat about the uncertainties in the interference of the two 
terms
discussed above,
we obtain reasonable production
rates in
spite of the fact that the momentum transfer is not too small.

%On the
%other
%hand, the spectra obtained, assuming a width of 7 MeV for the
%$\Theta^+$ nuclear
%states, or the experimental resolution should the width be even
%smaller, were
%clean enough to allow a separation of the different states, in the
%sense that
%the separation of levels is reasonably bigger than the width.

Measurements of binding energies and partial decay widths in nuclei 
would
provide precise information on the coupling of the $\Theta^+$ to two 
meson
channels and about the $K\pi N$ component in the $\Theta^+$ wave 
function.
The results obtained here should strongly encourage to do this 
experiment
which could open the doors
to the new field of $\Theta^+$ hypernuclei.

%\newpage

\section*{Acknowledgments}
This work is partly supported by DGICYT contract number BFM2003-00856,
and the E.U. EURIDICE network contract no. HPRN-CT-2002-00311.
This research is part of the EU Integrated Infrastructure Initiative
Hadron Physics Project under contract number RII3-CT-2004-506078.

%\section{Introduction}

%%%%%%%%%%%%%%%%%%%%%% Figure %%%%%%%%%%%%%%%%%%%%%%%%%
%%%%%%%%%%%%%%%%%%%%%%%%%%%%%%%%%%%%%%%%%%%%%%%%%%%%%%%%%%%%

%%%%%%%%%%%%%%%%%%%%%% Figure %%%%%%%%%%%%%%%%%%%%%%%%%
%%%%%%%%%%%%%%%%%%%%%%%%%%%%%%%%%%%%%%%%%%%%%%%%%%%%%%%%%%%%

%%%%%%%%%%%%%%%%%%%%%% Figure %%%%%%%%%%%%%%%%%%%%%%%%%
%%%%%%%%%%%%%%%%%%%%%%%%%%%%%%%%%%%%%%%%%%%%%%%%%%%%%%%%%%%%

\end{document}